# U-70 PROTON SYNCHROTRON EXTRACTED BEAM LINES CONTROL SYSTEM MODERNIZATION

V.Alferov, Y.Bordanovski, S.Klimov, V.Ilukin, V. Kuznetsov, O.Radin
A.Shalunov, A.Sytin, P.Vetrov, V.Yaryguine, V.Zapolsky, Zarucheisky
Institute for High Energy Physics, Protvino, Russia

Abstract

A 70 GeV Proton Synchrotron Extracted Beam Lines Control System is described. About 130 Magnet Dipoles and Quadrupoles, 20 Correction Magnets, 50 Beam Collimators, BPM equipment spread over 1 Km have to be controlled. The old System was based on the PDP-11/40 and LSI-11 compatible computers and the MIL 1553 STD as a Field Bus. It successfully operated about 15 years. A new system includes home made Equipment Controllers based on I 8051 Processors, CAN Field Bus, FECs, Servers, Consoles connected by Ethernet. On the first stage of modernization PDPs and LSIs are replaced with PCs connected by Ethernet. Equipment controllers are being successfully tested in the Collimators and Corrector Magnets Controls during a run.

## 1 INTRODUCTION

The Extracted Beams on the Serpukhov 70 GeV Proton Synchrotron are spread for over 1 km. They include about 130 Magnet Dipoles and Quadrupoles, 20 small Correction Magnets, 50 Beam Collimators, BPM, vacuum, interlock and other equipment. The Power Supplies (PS) of the Dipoles and Quadrupoles are installed in the special building 500m away. Total number of the I/O signals about 1500.

The old Control System was designed in the early 80-th [1]. It included PDP-11/40 compatible Host Computer, two LSI-11 compatible FECs, five MIL 1553 Field Buses to distribute the digital commands to the PSs as well as to connect the two Beam Control Rooms with the PS Building Control Room, Digital Voltmeters and distributed multiplexers for the control of the magnets current. During more than 10 years the System demonstrated high reliability but by time and because of heavy condition electronics as well as computers got old and lost their reliability. Its disadvantage also was slow (tenths of seconds) the setting of regimes. A parallel access to the equipment from different control rooms also was impossible.

A new project is based on the standard 3-layer model. The backbone of the system is an Ethernet LAN connecting a Power Supplies Building with the Beam Lines area including experimental control rooms. An upper level is presented by PCs, a middle level – by special Field Bus controllers GPFC [2], a lower level – by home made Equipment Controllers (EC). GPFCs and ECs will be implemented on the second stage of upgrading, after testing with a small group of equipment. On the first stage we implemented PCs connected by Ethernet without changing interface electronics.

## 2 SYSTEM ARCHITECTURE

The layout of the control system is shown in fig. 1. There are three main groups of controlled equipment. The most numerous are Dipoles and Quadrupoles PSs in the PSB. They are connected to PC in the PSB Control Room by four MIL 1553 Field Buses. A CAMAC Crate houses relevant Bus Controllers. Field Bus is used for distribution of regime digital commands to DACs of PSs. Control of regimes is provided by means of distributed Analog Multiplexer connecting PS Shunts to the Digital Voltmeter. A Status Register is used for controls of polarity, water cooling etc.

A second group of equipment are low current PSs of Correction Magnets distributed along Beam Lines. They are controlled from two Beam Control Rooms in the same way.

A third group are Beam Collimators. The multiplexed Collimator Controllers (CAMAC modules in two Beam Control Rooms) perform position control with help of shaft encoders and length modulation controls of DC Motors.

## 3 THE SOFTWARE DESCRIPTION

The software is developed under MS Windows 98/NT using Visual C++/Visual Basic, and client-server approach, based on DCOM (fig 2). The configuration of the PSs is placed in the Microsoft Access database.

The complex data exchange between different PC's in the different buildings, and user friendly interfaces are available. The powerful database tools provide different possibilities in the post-mortem analysis, easy configuration and extensibility of the data tables.

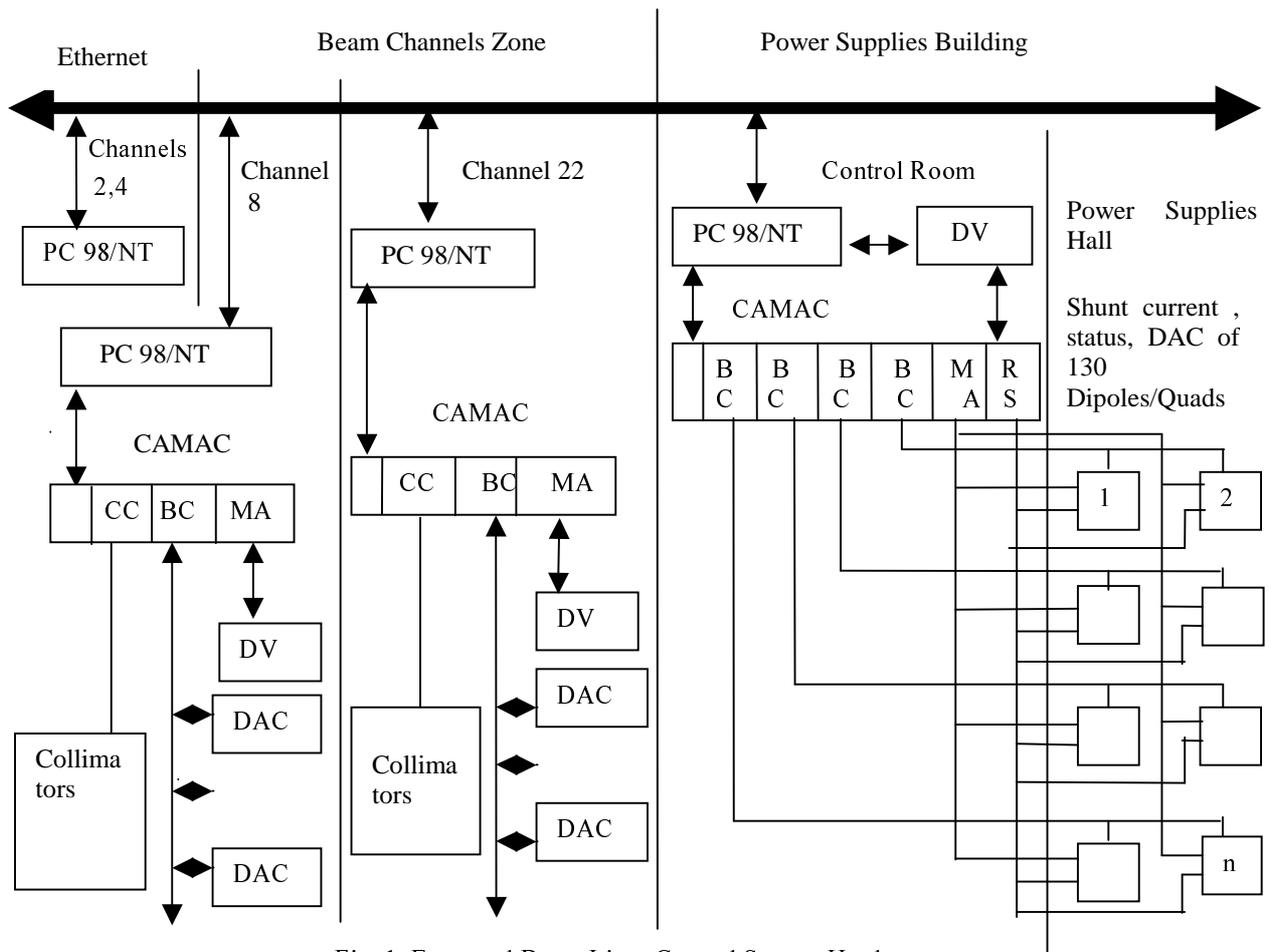

Fig .1. Extracted Beam Lines Control System Hardware
CC – Collimator Controller, BC – MIL 1553 BUS Controller, MA – Analog Multiplexer,
DAC – Digital to Analog Converter, DV – Digital Voltmeter, RS – Status Register.

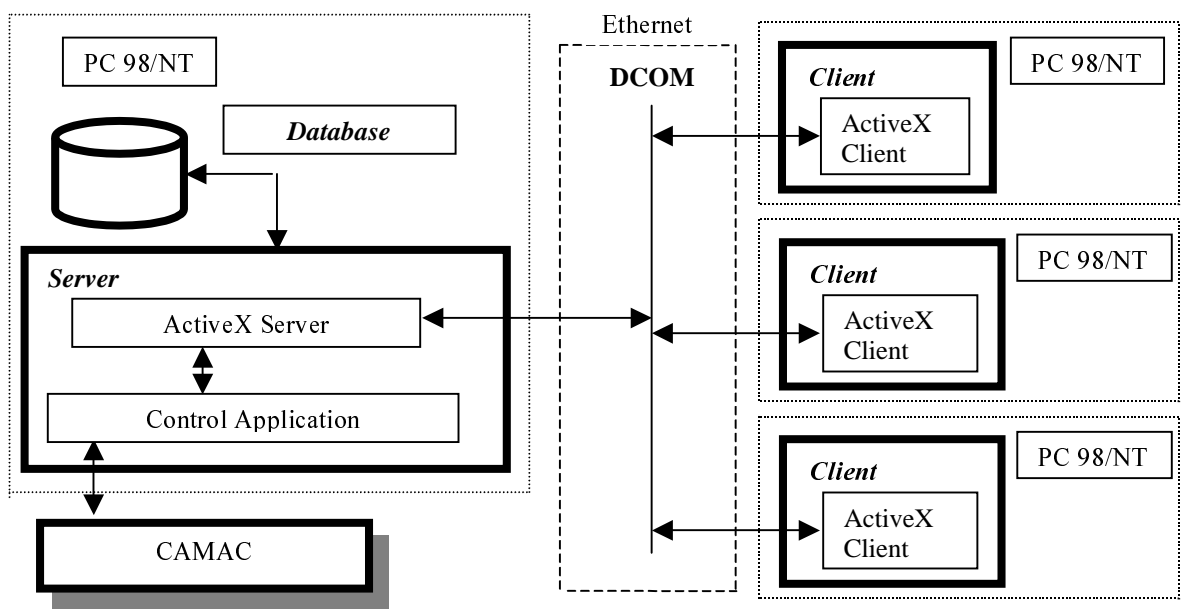

Figure 2: The structure of the Power Supplies control system software.

Control of the PSs from the different control rooms in more convenient way is also available, the conflicts with the simultaneous access to the equipment are down to the minimum.

The new features to access the data through the web browsers have been added.

The implementation of the equipment controllers on the second stage of the upgrade will provide a fully parallel access to the Power Supplies and Collimators.